\documentclass[11pt]{iopart}

\usepackage[top=30truemm,bottom=30truemm,left=25truemm,right=25truemm]{geometry}
\usepackage[dvipdfmx]{graphicx}
\usepackage{epstopdf}
\usepackage{float}
\usepackage{cite}

\usepackage{dcolumn} 
\usepackage{bm}
\usepackage{color}

\begin{document}
\def\ii{{\mathrm{i}}}
\def\ee{{\mathrm{e}}}
\def\dd{{\mathrm{d}}}
\def\Re{\mathop{\mathrm{Re}}}
\def\Im{\mathop{\mathrm{Im}}}

\def\bra#1{\langle #1|}
\def\ket#1{|#1\rangle}
\def\bracket#1{\langle#1\rangle}

\def\sub#1{_{\mathrm{#1}}}
\def\sur#1{^{\mathrm{#1}}}
\def\vct#1{{\mathchoice{\mbox{\boldmath$#1$}}{\mbox{\boldmath$#1$}}%
  {\mbox{\scriptsize\boldmath$#1$}}{\mbox{\scriptsize\boldmath$#1$}}}}

\newcommand{\nn}{\nonumber \\}

\title[Boltzmann sampling for an XY model using an NOPO network]
{Boltzmann sampling for an XY model using a non-degenerate optical parametric oscillator network}
\author{Y Takeda$^{1,2}$, S Tamate$^2$, Y Yamamoto$^{3,4}$, H Takesue$^5$, T Inagaki$^5$ and S Utsunomiya$^{1,2}$}

\address{$^1$ Department of Physics, Faculty of Science, Tokyo University of Science, 1-3 Kagurazaka, Shinjuku-ku, Tokyo 162-8601, Japan}
\address{$^2$ National Institute of Informatics, Hitotsubashi 2-1-2, Chiyoda-ku, Tokyo 101-8403, Japan}
\address{$^3$ ImPACT Program, Japan Science and Technology Agency, Gobancho 7, Chiyoda-ku, Tokyo 102-0076, Japan}
\address{$^4$ E. L. Ginzton Laboratory, Stanford University, Stanford, CA 94305, USA}
\address{$^5$ NTT Basic Research Laboratories, NTT Corporation, 3-1 Morinosato Wakamiya, Atsugi, Kanagawa 243-0198, Japan}

\begin{abstract}
We present an experimental scheme of implementing multiple spins in a classical XY model using a non-degenerate optical parametric oscillator (NOPO) network. We built an NOPO network to simulate a one-dimensional XY Hamiltonian with 5000 spins and externally controllable effective temperatures. The XY spin variables in our scheme are mapped onto the phases of multiple NOPO pulses in a single ring cavity and interactions between XY spins are implemented by mutual injections between NOPOs. We show the steady-state distribution of optical phases of such NOPO pulses is equivalent to the Boltzmann distribution of the corresponding XY model. Estimated effective temperatures converged to the setting values, and the estimated temperatures and the mean energy exhibited good agreement with the numerical simulations of the Langevin dynamics of NOPO phases.
\end{abstract}
\maketitle

\section{Introduction}
The dynamics of various spin models, such as the Ising and XY models, have been studied extensively in statistical mechanics. While simulating these dynamics is useful for understanding magnetism and spin glasses, they can be utilized for various network-based approaches to computation\cite{Ross1980,Ackley1985,Baldi1990,Zemel1995}. Generating a large number of different states according to a Boltzmann distribution, is not only required for calculating the expectation values of various observables in these models, but also in a wide range of applications, e.g., computer vision\cite{Li2009} and reinforcement learning\cite{Sallans2004}. The Markov chain Monte Carlo (MCMC) method is commonly employed to achieve this task. However, the spin systems in real applications are often frustrated and rugged in energy landscapes, so that the sampling procedures to generate multiple independent states suffer from slow relaxation of the Markov chains. Various methods have been proposed to enhance the relaxation process of the MCMC sampling technique, such as exchange Monte Carlo, simulated tempering and multi-canonical Monte Carlo\cite{Iba2000}. 

Alternative approaches to sampling these models have recently been investigated. Particular physical systems such as laser networks\cite{Utsunomiya2011} and superconducting circuits\cite{Johnson2011} have been designed to implement required Hamiltonians so that multiple samples can be drawn through iterative measurements of states because of their stochastic dynamics. If the probability distribution of generated spin configurations approximately follows Boltzmann statistics, computationally hard tasks involving Boltzmann sampling become feasible\cite{Sakaguchi2016,Crawford2016,Dumoulin2013}. Temperature parameters also need to be tuned to implement arbitrary Boltzmann distributions.

The XY model is a classic spin model in statistical mechanics in which an XY spin has a continuous direction in a two-dimensional plane. Thus, XY spins can be associated with a specific type of data that is called directional \cite{Mardia2000}. Various probabilistic models involving directional variables have been proposed to analyze real-world directional data such as torsion angles in biomolecules\cite{Razavian2011,Mardia2007,Mardia2008,Hamelryck2012}. The XY model can be simulated by using optical cavity systems. Experiments using a coupled laser system\cite{Tamate2016,Nixon2013} and a coupled polariton system\cite{Berloff2016} have recently been reported. A scheme using a network of NOPOs has been studied through numerical simulations\cite{Hamerly2016}.

This paper describes how we implement an NOPO network to simulate a one-dimensional classical XY model. Highly-nonlinear-fiber-based and periodically poled lithium niobate (PPLN)-based OPO networks have recently been experimentally studied to implement Ising Hamiltonians using degenerate optical parametric oscillators (DOPOs)\cite{Inagaki2016a,Inagaki2016b,McMahon2016}. We employed an experimental scheme similar to that used in \cite{Inagaki2016a} but with important modification from DOPO to NOPO to implement XY spins. Our system features a large number of XY spins ($N$ = 5000) and a configureable temperature parameter. 

\section{Theoretical background}
\label{sec:2}
\subsection{Hamiltonian}
\label{sec:theory}
XY spin variables are expressed as unit-length vectors, $\mathbf{s}_i = (\cos\theta_i,\sin\theta_i)$, where $\theta_i \in [-\pi,\pi)$. The Hamiltonian of the system which comprises $N$ spins with configuration $\vct{\theta} = (\theta_1,...,\theta_N)$, is given as:
\begin{equation}
	\mathcal{H}(\vct{\theta}) = -\sum^N_{k,l:k<l}J_{kl}~\mathbf{s}_k\cdot \mathbf{s}_l = -\sum^N_{k,l:k<l}J_{kl} \cos(\theta_k-\theta_l) \label{eq:xyh},
\end{equation}
where $J_{kl}$ is the interaction strength between the $k$-th and $l$-th spin. The XY model is implemented using an NOPO network in which part of the signal field of one NOPO is coherently injected into other NOPOs. The signal field in an NOPO takes an arbitrary phase, so that an XY spin can be mapped onto the optical phase of an NOPO. The phase dynamics of an NOPO network are governed by the two counteracting forces: drift force that reduces the phase difference between connected NOPOs and diffusion force that randomly fluctuates the phase. The steady state distribution of these phases is governed by the balance between the two forces.

\subsection{Optical parametric oscillation}
The NOPO is obtained by pump-degenerate optical parametric amplification based on four-wave mixing (FWM), where the two pump waves are degenerate and the signal and idler waves are non-degenerate.
Their frequencies  satisfy $2f_\mathrm{p} = f_\mathrm{s} +  f_\mathrm{i}$, where $f_\mathrm{p},f_\mathrm{s}$ and $f_\mathrm{i}$ are the frequencies of the degenerate pump, signal and idler respectively.
Let us consider a case where the signal, idler, and pump fields are confined in a ring cavity.
The classical amplitude of the signal, idler, and pump fields are denoted by $a\sub{s}$, $a\sub{i}$, and $a\sub{p}$.
These three fields interact due to the $\chi^{(3)}$ nonlinearity in an optical fiber.
The $\chi^{(3)}$ nonlinearity gives rise to self-phase modulation (SPM), cross-phase modulation (XPM) and FWM. Here, let us consider pump-degenerate FWM.
We assumed that the effects of SPM and XPM would become negligibly small by selecting appropriate phase matching conditions.
We can only leave the term of FWM in such a case, and the equations of motion for three fields are given by:
\begin{eqnarray}
  \frac{\dd}{\dd t} a\sub{p} &= -\frac{\gamma\sub{p}}{2}a\sub{p} - \kappa a\sub{p}^*a\sub{s}a\sub{i}
  + \sqrt{\gamma\sub{p}}F\sub{p}  \label{eq:1} \\
  \frac{\dd}{\dd t} a\sub{s} &=
  -\frac{\gamma\sub{s}}{2}a\sub{s}
  + \frac{\kappa}{2} a\sub{i}^*a\sub{p}^2  \label{eq:2} \\
  \frac{\dd}{\dd t} a\sub{i} &=
  -\frac{\gamma\sub{i}}{2}a\sub{i}
  + \frac{\kappa}{2} a\sub{s}^*a\sub{p}^2,  \label{eq:3}
\end{eqnarray}
where $\kappa$ denotes the strength of parametric coupling. The cavity decay rates for the signal, idler, and pump fields are denoted by $\gamma\sub{s}$, $\gamma\sub{i}$, and $\gamma\sub{p}$. The $F\sub{p}$ denotes the amplitude of an external pump field.

Here, let us consider a situation where the cavity decay rates of the pump and idler fields are much larger than that of the signal field: $\gamma\sub{s} \ll \gamma\sub{p}, \gamma\sub{i}$. Adiabatically eliminating the pump and idler fields by assuming $\dd a\sub{p}/\dd t = 0$ and $\dd a\sub{i} /\dd t = 0$, we can obtain the equation of motion for the signal field:
\begin{eqnarray}
  \frac{\dd}{\dd t} a\sub{s}
  = \frac{\gamma\sub{s}}{2}\left[
  \left\{ s(|a\sub{s}|^2 / n_0) \frac{|F\sub{p}|}{F\sub{p}\sur{(th)}}
  \right\}^4
  - 1 \right]
  a\sub{s}.  \label{eq:one_pump_opo},
\end{eqnarray}
where $s(x)$ is a parameter representing the gain saturation. The saturation photon number is represented by $n_0$. The amplitude of the external pump field at the threshold is denoted by $F\sub{p}\sur{(th)}$. The explicit forms of $s(x)$, $n_0$ and $F\sub{p}\sur{(th)}$ are given as:
\begin{eqnarray}
  s(x) &=
  \left[
    1 + \left(
      \sqrt[3]{-\frac{\sqrt{x}}{2} + \sqrt{\frac{x}{4} + \frac{1}{27}}}
      - \sqrt[3]{\frac{\sqrt{x}}{2} + \sqrt{\frac{x}{4} + \frac{1}{27}}}
    \right)^2
  \right]^{-1},  \label{eq:5} \\
  n_0 &= \frac{\gamma\sub{p}^2\gamma\sub{i}}{8\kappa^2 |F\sub{p}|^2},\quad
  F\sub{p}\sur{(th)} = \left( \frac{\gamma\sub{p}\sqrt{\gamma\sub{s}\gamma\sub{i}}}{2\kappa}\right)^{1/2}.  \label{eq:6}
\end{eqnarray}

Note that the gain term in \eref{eq:one_pump_opo} is insensitive to the phase of the signal field. As a result, when the external pump field is above the threshold, i.e., $|F\sub{p}| > F\sub{p}\sur{(th)}$, the signal field oscillates in an arbitrary phase. The steady state photon number of the signal field is given by:
\begin{eqnarray}
  n\sur{(ss)} = n_0\left[
    \left(\frac{|F\sub{p}|}{F\sub{p}\sur{(th)}} - 1\right)^{3/2} + \left(\frac{|F\sub{p}|}{F\sub{p}\sur{(th)}} - 1\right)^{1/2}
  \right]^2,  \label{eq:7}
\end{eqnarray}
and the amplitude of the signal field can be expressed as $a\sub{s} = \sqrt{n\sur{(ss)}}\ee^{i\theta}$ with an arbitrary phase. We can use this $U(1)$-degree of freedom of the NOPO signal field as an XY spin.

\subsection{Langevin equation for NOPO network}
Suppose that multiple NOPOs with the same properties are mutually connected with mutual optical injection. In addition to the optical connection, we assume that each NOPO field is driven by white noise, which can be intrinsic vacuum fluctuations or excess classical noise.
If we denote the signal field of $k$-th NOPO as $a_k$, the Langevin equation for the NOPO network can be written as:
\begin{eqnarray}
  \frac{\dd}{\dd t} a_k
  = \frac{\gamma\sub{s}}{2}\left[
  \left\{ s(|a_k|^2 / n_0) \frac{|F\sub{p}|}{F\sub{p}\sur{(th)}}
  \right\}^4
  - 1 \right]
  a_k + \frac{\gamma\sub{inj}}{2} \sum_{l: l\neq k}J_{kl} a_l
  + \sqrt{D} \xi_k(t).  \label{eq:8}
\end{eqnarray}
The mutual injection rate is denoted by $\gamma\sub{inj}$ and the connectivity of the NOPO network is represented by a matrix, $J_{kl}$. The complex-valued noise function, $\xi_k(t)$, is assumed to be $\delta$-correlated: $\bracket{\xi_k(t)} = 0, \bracket{\xi_k^*(t) \xi_l(t')} = 2\delta_{kl}\delta(t-t')$ and $\bracket{\xi_k(t) \xi_l(t')} = 0$. The diffusion coefficient is represented by $D$.

If we denote the photon number and phase of the $k$-th NOPO field as $n_k$ and $\theta_k$, we can separate the Langevin dynamics into the photon number and the phase parts as:
\begin{eqnarray}
  \frac{\dd n_k}{\dd t} =& \gamma\sub{s}\left[
  \left\{ s(n_k / n_0) \frac{|F\sub{p}|}{F\sub{p}\sur{(th)}}
  \right\}^4
  - 1 \right] n_k \nn
  &+ \gamma\sub{inj}\sqrt{n_k n_l}\sum_{l:l\neq k}J_{kl}\cos(\theta_k - \theta_l) + 2D  + 2\sqrt{D n_k} \xi^\mathrm{(n)}_k(t),  \label{eq:9} \\
  \frac{\dd \theta_k}{\dd t} =&
  - \frac{\gamma\sub{inj}}{2} \sum_{l:l\neq k} \sqrt{\frac{n_l}{n_k}} J_{kl}\sin(\theta_k - \theta_l)  + \sqrt{\frac{D}{n_k}} \xi^{(\theta)}_k(t),\label{eq:10},
\end{eqnarray}
where the real-valued noise functions, $\xi^\mathrm{(n)}_k(t)$ and $\xi^{(\theta)}_k(t)$, satisfy $\bracket{\xi^{(x)}_k(t)\xi^{(y)}_l(t')} = \delta_{xy}\delta_{kl}\delta(t-t')$, where $x,y=\mathrm{n},\theta$ and $k,l=1,...,N$.

Let us consider the case when $\gamma\sub{s} \gg \gamma\sub{inj}$; the photon number dynamics are much faster than the phase dynamics. Assume that the mutual injection is sufficiently small so that the steady state photon numbers of all NOPOs are identical and denoted by $n\sur{(ss)}$. The Langevin equation for the NOPO phases reduces to:
\begin{eqnarray}
  \frac{\dd \theta_k}{\dd t} = - \frac{\gamma\sub{inj}}{2}\sum_{l:l\neq k} J_{kl}\sin(\theta_k - \theta_l) + \sqrt{D_\theta} \xi^{(\theta)}_k(t), \label{eq:Kuramoto}
\end{eqnarray}
where $D_\theta = D/n\sur{(ss)}$ represents the phase diffusion coefficient. \Eref{eq:Kuramoto} is known as the Kuramoto model driven by noise \cite{Acebron2005}.
The drift term in \eref{eq:Kuramoto} can be expressed by using the potential function as:
\begin{eqnarray}
  \frac{\dd \theta_k}{\dd t} &= - \frac{\gamma\sub{inj}}{2} \frac{\partial H(\vct{\theta})}{\partial \theta_k} + \sqrt{D_\theta}\xi^{(\theta)}_k(t),  \label{eq:potential_form} \\
  H(\vct{\theta}) &= - \sum_{k,l: k < l} J_{kl}\cos(\theta_k - \theta_l),
\end{eqnarray}
where the potential function has the same form as the classical XY spin model. The steady-state probability distribution for the phase configuration $\vct{\theta}$ is given by the following Gibbs-Boltzmann distribution \cite{Risken1974}:
\begin{eqnarray}
  P(\vct{\theta}) &= \exp \left( -\beta H(\vct{\theta}) \right), \\
  \beta &= \frac{\gamma\sub{inj}}{D_\theta}. \label{eq:beta}
\end{eqnarray}
Thus, we find that the thermal equilibrium state of the classical XY model is realized as the steady-state distribution of the mutually coupled NOPO network. Note that the effective temperature of the simulated XY model can be tuned by changing the ratio between the injection rate $\gamma\sub{inj}$ and the phase diffusion coefficient $D_\theta$.

\section{Experimental setup}
\label{sec:setup}

\begin{figure}
\begin{center}
\includegraphics{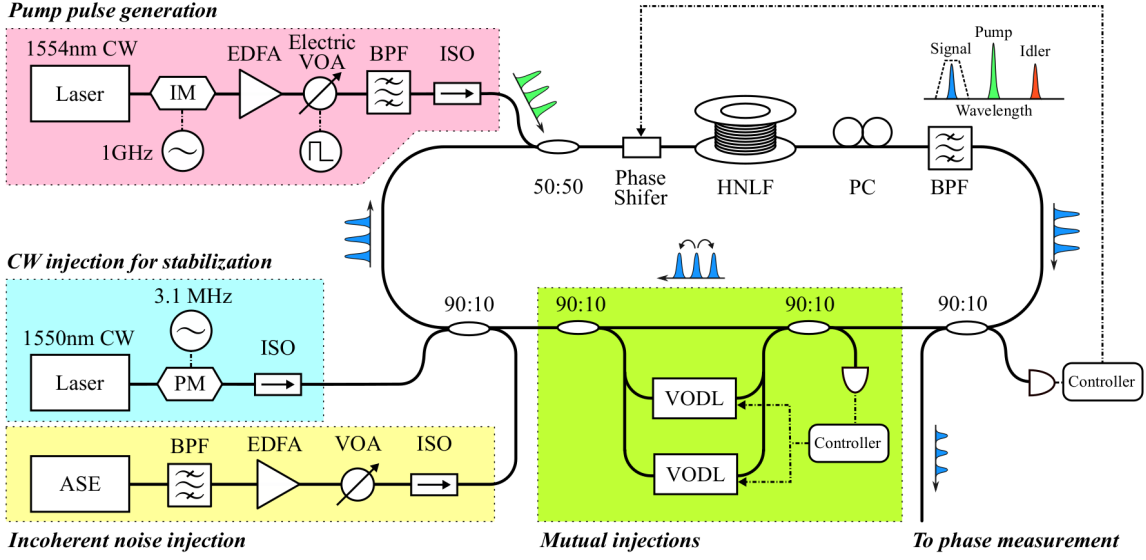}
\caption{A schematic of experimental setup. The pump laser field (1554 nm) is externally intensity-modulated into pulses with 26.2-ps full width at half maximum (FWHM) and 1-GHz repetition rate. The pump pulses are injected into the main fiber cavity through a 50:50 optical coupler. The signal pulse (1550 nm) is produced by synchronous pumping. The FWHM temporal width of the signal pulses is 32.6 ps and the FWHM bandwidth is $\sim$0.1nm. The cavity length is $\sim$1-km, so that the cavity supports 5000 NOPO pulses. Mutual injections between NOPOs are achieved with two optical delay lines; the first is for injection to the forward NOPO pulses and the second is for the backward NOPO pulses. The signal field is extracted with a 90:10 coupler for phase measurements. Incoherent noise (center wavelength: 1550 nm, FWHM bandwidth: $\sim$0.1 nm) is injected through a 90:10 coupler to control the phase diffusion coefficient. While the NOPO pulses propagate clockwise, an external cw laser (1550 nm) is injected counterclockwise to stabilize the lengths of the cavity and two delay lines by using feedback control to the phase shifters. HNLF: highly nonlinear fiber (nonlinear parameter $\gamma = 21~\mathrm{W^{-1}\cdot km^{-1}}$, group dispersion parameter $\beta_2 = -0.31~\mathrm{ps^2\cdot km^{-1}}$ at 1550 nm and length $L = 930~\mathrm{m}$), VOA: variable optical attenuator, VODL: variable optical delay line (containing VOA to adjust the injection rate $\gamma_\mathrm{inj}$), IM: intensity modulator, PM: phase modulator, EDFA: erbium-doped fiber amplifier, BPF: optical band pass filter, ISO: isolator, ASE: amplified spontaneous emission source, and PC: polarization controller.}
\label{figure1}
\end{center}
\end{figure}

We employed the time-devision multiplexing (TDM) method to connect a large number of NOPOs mutually. The multiple NOPOs are generated as optical pluses in a single optical ring cavity. The signal fields of different NOPOs are mutually injected with optical delay lines to implement the interactions between XY spins.
The interaction coefficient of an XY model can be independently set by the intensities and phases of mutual injections.

Figure \ref{figure1} is the schematic for our experimental setup.
We used highly nonlinear fiber (HNLF) as an optical four-wave mixer.
The pump and idler waves are attenuated by an optical bandpass filter at the output of HNLF, so that only amplified signal pulses stay inside a cavity.
We implemented a ferromagnetic one-dimensional XY model in this study.
There are optical delay lines with $\pm 1$-interval delays to accomplish one-dimensional nearest neighbor couplings, where both the intensity and phase of mutual injections are fixed for all connections.
Thus the implemented Hamiltonian is $\mathcal{H}(\vct{\theta})=-\sum_{k=1}^N \cos(\theta_{k+1}-\theta_k)$, where $\theta_{N+1} = \theta_{1}$ and $N=5000$. The relative phases of individual NOPOs to the external cw field of the 1550-nm wavelength were measured.
We also measured the relative phases between two adjacent NOPO pulses. (The phase measurement techniques used in this study are described in \ref{sec:PM}.)

In our scheme, the effective temperature $\beta\sur{(eff)}$ can be controlled by the injection rate $\gamma_\mathrm{inj}$ and the phase diffusion coefficient $D_\theta$ as expressed in \eref{eq:beta}. The configurable range of $\beta\sur{(eff)}$ is limited by the experimental constraints of $\gamma_\mathrm{inj}$ and $D_\theta$. The phase diffusion coefficient $D_\theta$ is controlled by the external noise injection power.
We injected incoherent cw field at the signal wavelength through a coupler to increase phase noise in addition to the intrinsic phase noise in the NOPOs. The lower bound of $D_\theta$ is $\sim$0.44 kHz in this system, which was determined by measurering the intrinsic phase diffusion noise in this optical cavity.
The injection rate $\gamma_\mathrm{inj}$, on the other hand, can be configured with the transmittance $T$ of $\pm 1$-interval optical delay lines as $\gamma\sub{inj} = 2\sqrt{T}/\tau\sub{rt}$, where $\tau\sub{rt} = 5~\mathrm{\mu s}$ is the cavity round trip time. The maximum possible value of $\gamma\sub{inj}$ is $\sim$15 kHz. (A case in which $\beta\sur{(set)}$ is configured with $\gamma_\mathrm{inj}$ is discussed in \ref{sec:gamma_inj}.)

\section{Results}
\subsection{Oscillations of NOPO signal field}
\begin{figure}[htbp]
 	\begin{minipage}{0.31\hsize}
  		\begin{center}
   		\includegraphics{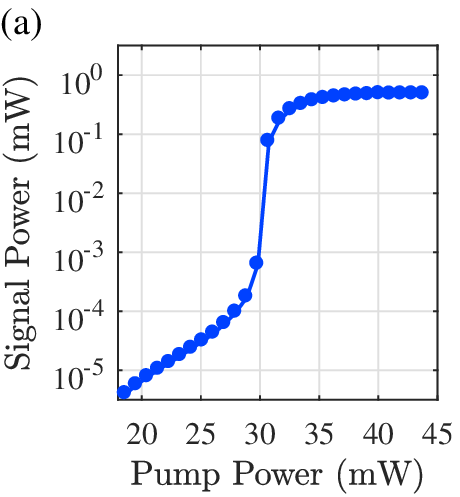}
  		\end{center}
 	\end{minipage}
	\begin{minipage}{0.37\hsize}
  		\begin{center}
   		\includegraphics{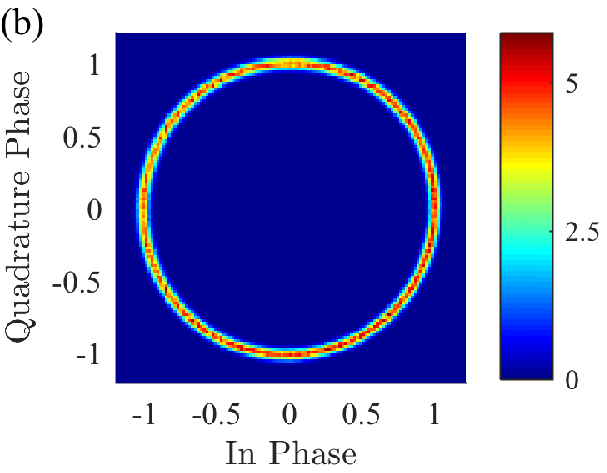}
  		\end{center}
 	\end{minipage}
 	\begin{minipage}{0.3\hsize}
  		\begin{center}
   		\includegraphics{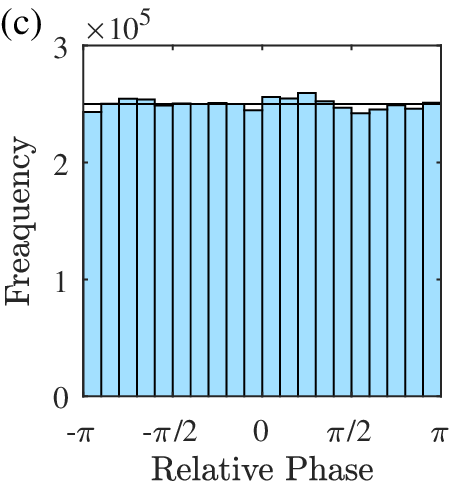}
  		\end{center}
 	\end{minipage}
\caption{(a) An output power of NOPO signal field. The oscillation threshold pump power is $\sim$30 mW. Well above the threshold, the intracavity signal power is $\sim$480 $\mathrm{\mu W}$. (b) 2D histogram of observed in-phase/quadrature-phase amplitudes of the one-interval delay interference signal, without mutual injections. Each point is regarded as a phasor of a relative phase angle between two adjacent NOPO pulses. (c) Histogram of relative phase angles extracted from IQ signals in (b). The black solid line at $2.5\times10^5$ indicates the expected value for each bin when a uniform distribution is assumed. This relative phase distribution is almost uniform so that the NOPO signal phases can be regarded as being independent.}
\label{figure2}
\end{figure}

\Fref{figure2}(a) plots the output power of the signal field for various pump powers. Clear threshold behavior was observed at a pump power of 30 mW. We then experimentally confirmed that the phases of multiple NOPO pulses inside the same fiber cavity are mutually independent when there were no mutual injections.
We physically blocked the delay lines and acquired relative phases repeatedly 100 ms after the pump was switched on. The number of acquired samples was 1000 (One sample contained 5000 relative phase values). \Fref{figure2} (b) has a 2D histogram of in-phase/quadrature-phase (IQ) signals obtained from a one-interval delay interferometer. Each data point can be regarded as a phasor of a relative phase angle between two adjacent NOPO pulses. \Fref{figure2} (c) is a histogram of relative phases. The solid line at $2.5\times10^5$ is the expected value for one bin. As the distribution we obtained is almost uniform, each of the NOPO phases can be regarded as being independent.

\subsection{Control of the phase diffusion coefficient}
\begin{figure}[htbp]
	\begin{minipage}{0.66\hsize}
  		\begin{center}
   		\includegraphics{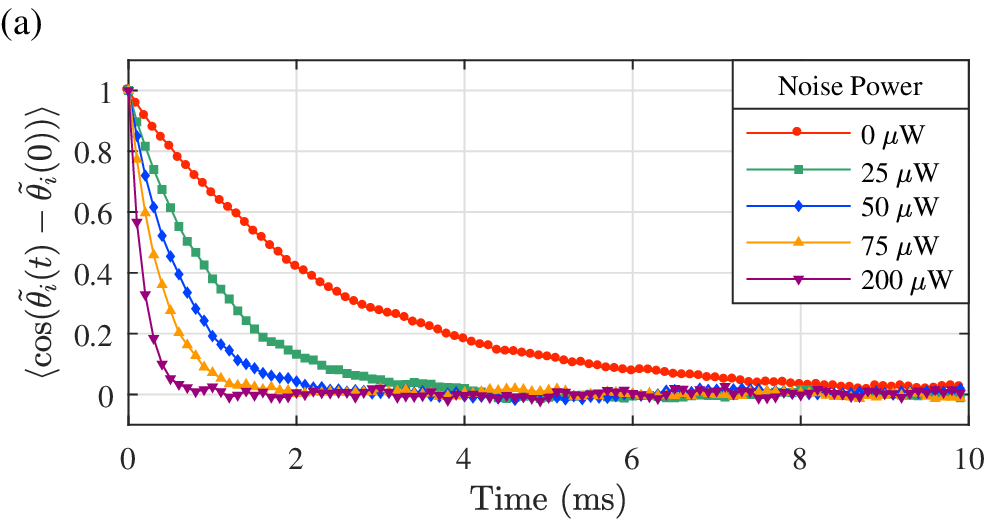}
  		\end{center}
 	\end{minipage}
 	\begin{minipage}{0.34\hsize}
  		\begin{center}
   		\includegraphics{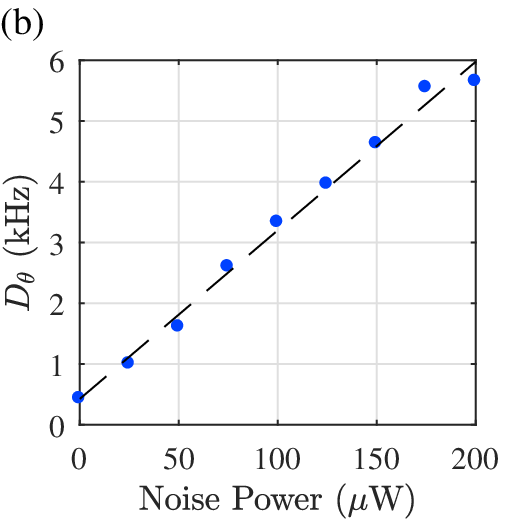}
  		\end{center}
 	\end{minipage}
\caption{(a) Phase correlation decay of NOPOs. Phase diffusion coefficient $D_\theta$ can be estimated simply by fitting these decay curves with $\exp(-D_\theta t)$. (b) Estimated phase diffusion coefficients $D_\theta$ for various injection powers of external incoherent noise. The dashed line plots simple linear regression with the $D_\theta$ intercept fixed to the data point. The finite $D_\theta$ value at noise power zero determines the upper bound for possible temperature parameter $\beta\sur{(set)}$ that was configured in our system.}
\label{figure3}
\end{figure}

We found that phase diffusion could be increased by external noise injection. When an NOPO phase diffuses as $\mathrm{d}\theta_k/\mathrm{d}t= \sqrt{D_\theta}\xi^{(\theta)}_k(t)$, where $\bracket{\xi^{(\theta)}_k(t)\xi^{(\theta)}_l(t')} = \delta_{kl}\delta(t-t')$, 
the diffusion coefficient, $D_\theta$, can be readily estimated by measuring the decay rates of the cosines of relative phases:
\begin{equation}
	\langle \cos(\tilde{\theta}_k(t) - \tilde{\theta}_k(0)) \rangle =\exp(-D_\theta t), \label{eq:cor}
\end{equation}
where $\tilde{\theta}_k = \theta_{k+1} - \theta_{k}$. Figure \ref{figure3}(a) plots the measured values of \eref{eq:cor} for various injection powers of the external incoherent field. Phase diffusion coefficient $D_\theta$ for each condition is plotted in Figure \ref{figure3}(b), which was estimated by fitting the above decay trace of the cosines with $\exp(-D_\theta t)$. It can be seen that the phase diffusion coefficient $D_\theta$ is linearly increased with the injection power of external noise. 

\subsection{Control of the effective temperature}
We evaluated the controllability of the temperature parameter. First we measured relative phase distributions at different data acquisition times $t\sub{a}$ after the pump field was switched on, while we fixed the temperature parameter, $\beta\sur{(set)}= \gamma\sub{inj}/D_\theta$. We configured $\beta\sur{(set)} = 31$ and measured the relative phase distributions at $t_\mathrm{a}$ = 1, 10, 100 and 1000 ms. We acquired 1000 samples at $t_\mathrm{a}=$ 1-100 ms, and another 100 samples at 1000 ms.

\begin{figure}[htbp]
 	\begin{minipage}{0.5\hsize}
  		\begin{center}
   		\includegraphics{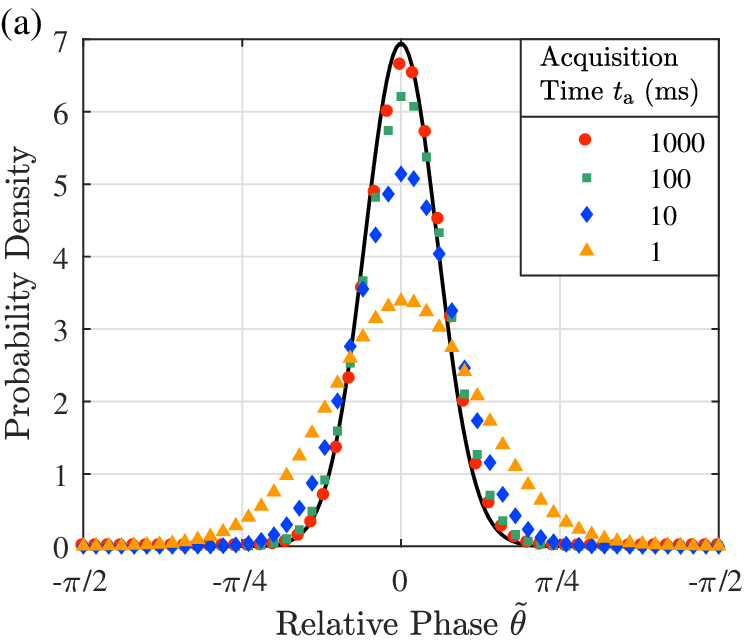}
  		\end{center}
 	\end{minipage}
 	\begin{minipage}{0.5\hsize}
  		\begin{center}
   		\includegraphics{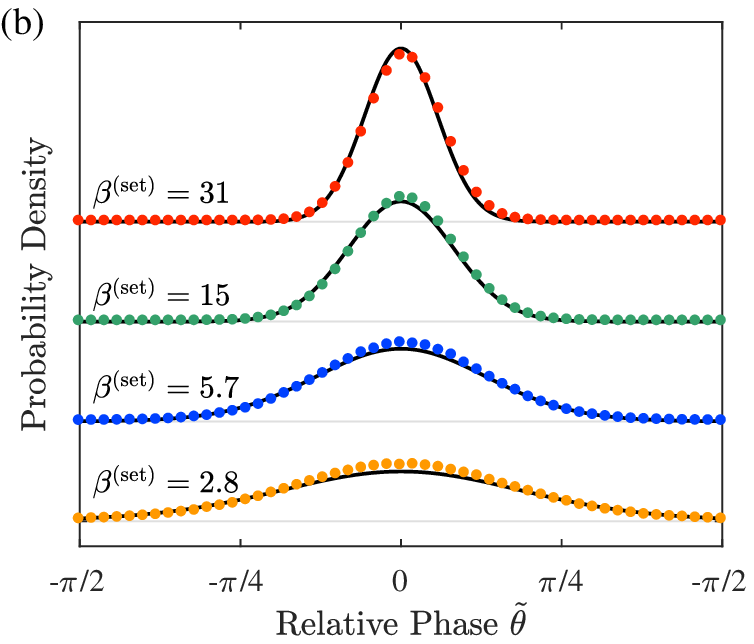}
  		\end{center}
 	\end{minipage}
\caption{Relative phase distributions. The markers indicate the data points obtained from the experiment. The black solid lines indicate the theoretical probability distributions \eref{eq:18} for each temperature parameter. (a) Case in which temperature parameter is fixed at $\beta\sur{(set)}=31$ and NOPO phases were acquired at different acquisition times, $t_\mathrm{a} = $1, 10, 100 and 1000 ms. (b) Relative phase distributions observed for different temperature parameters, $\beta\sur{(set)}$, at a same acquisition time, $t_\mathrm{a} = 1000$ ms. We acquired 1000 samples for $t_\mathrm{a}=$ 1-100 ms and acquired 100 samples for 1000 ms.}
\label{figure4}
\end{figure}

The probability distribution of the relative phase between adjacent XY spins in the 1D XY ring was calculated analytically (see \ref{sec:theoryxy}). It can be approximated in the case of a large number of spins as:
\begin{equation}
	p(\tilde{\theta}) = \frac{\exp (\beta \cos \tilde{\theta})}{2\pi I_0(\beta)}. \label{eq:18}
\end{equation}
where $I_n$ is the modified Bessel function of the first kind. The observed relative phase distributions are shown in \fref{figure4}(a). The black solid curve indicates the theoretical probability distribution \eref{eq:18} for $\beta = 31$. The relative phases concentrated as we took a longer time, and their distribution converged on the thermal equilibrium distribution for $\beta\sur{(set)}$.

We then fixed the data acquisition time, $t\sub{a} =$ 1000 ms, and measured the relative phase distributions for different temperature parameter, $\beta\sur{(set)} = 31,15,5.7$ and $2.8$. \Fref{figure4}(b) plots the relative phase distributions for each temperature parameter. The markers indicate the observed distributions and the solid curves show the theoretical probability distributions \eref{eq:18} for each temperature parameter. The concentration of the relative phases is controlled by temperature parameter $\beta\sur{(set)}$, and the distributions tended to agree with the theoretical curves for each $\beta\sur{(set)}$.

\begin{figure}[htbp]
 	\begin{minipage}{0.5\hsize}
  		\begin{center}
   		\includegraphics{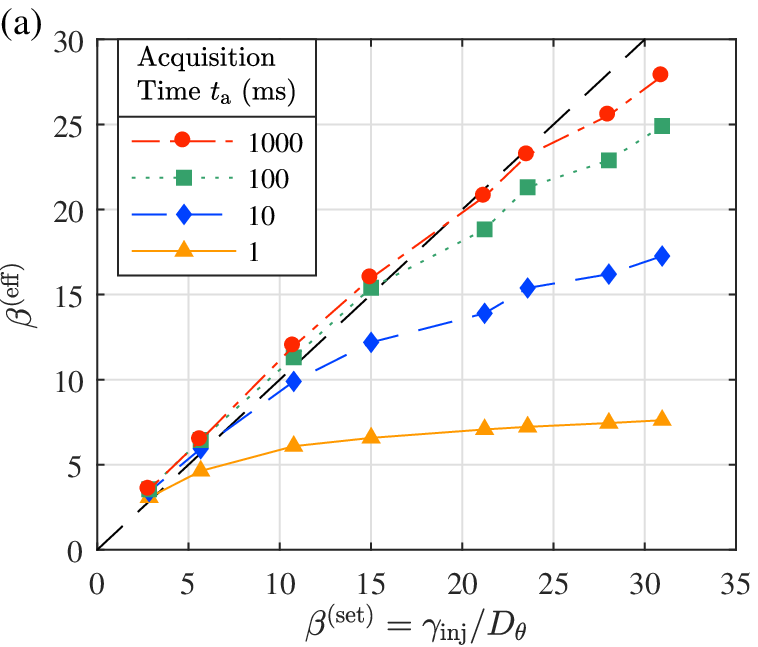}
  		\end{center}
 	\end{minipage}
 	\begin{minipage}{0.5\hsize}
  		\begin{center}
   		\includegraphics{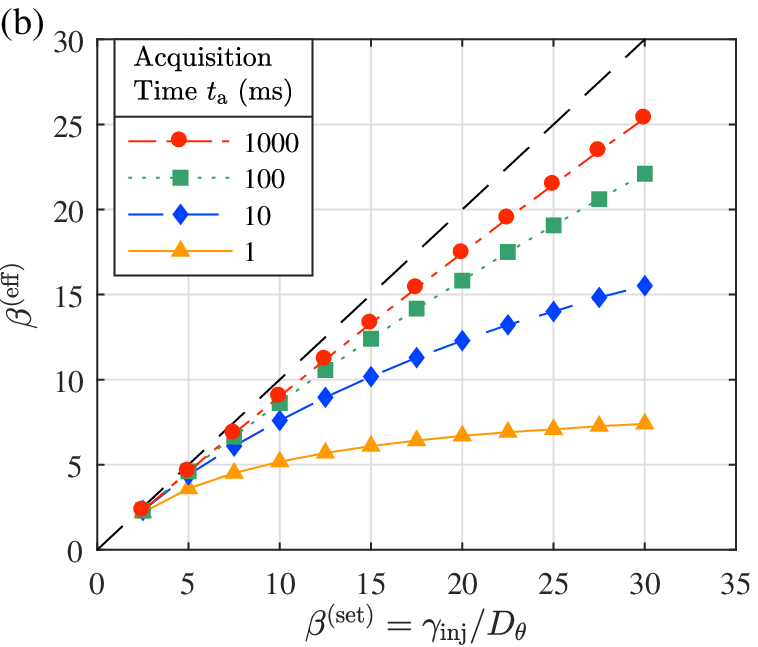}
  		\end{center}
 	\end{minipage}
\caption{Achieved effective temperature $\beta\sur{(eff)}$ for various settings $\beta\sur{(set)} = \gamma_\mathrm{inj}/D_\theta$. The $\beta\sur{(eff)}$ was estimated from the relative phase distribution. The setting values $\beta\sur{(set)}$ were varied by changing $D_\theta$, while $\gamma_\mathrm{inj}$ was fixed. (a) $\beta\sur{(eff)}$ values obtained from experiment. (b) $\beta\sur{(eff)}$ values estimated through numerical simulations. Black dashed line indicates where $\beta\sur{(eff)} = \beta\sur{(set)}$.}
\label{figure5}
\end{figure}

We repeatedly measured NOPO phases at different temperature parameter $\beta\sur{(set)}$ and different acquisition time $t_\mathrm{a}$, in order to comprehensively study the convergence of an effective temperature. We estimated the effective temperature $\beta\sur{(eff)}$ by fitting the acquired relative phase distribution with \eref{eq:18}.
\Fref{figure5}(a) plots the experimental effective temperature $\beta\sur{(eff)}$ for different $\beta\sur{(set)}$, in which the phases were measured at $t_\mathrm{a}$ = 1, 10, 100 and 1000 ms.
The black dashed line indicates where the effective temperature $\beta\sur{(eff)}$ is equal to its setting value $\beta\sur{(set)}$.
The effective temperature $\beta\sur{(eff)}$ came close to each setting value for later acquisition time $t_\mathrm{a}$ as seen in \fref{figure4}(a).
The time scale of convergence to the setting values became longer for high $\beta\sur{(set)}$ (i.e., low temperature configuration). 

\begin{figure}[htbp]
 	\begin{minipage}{0.5\hsize}
  		\begin{center}
   		\includegraphics{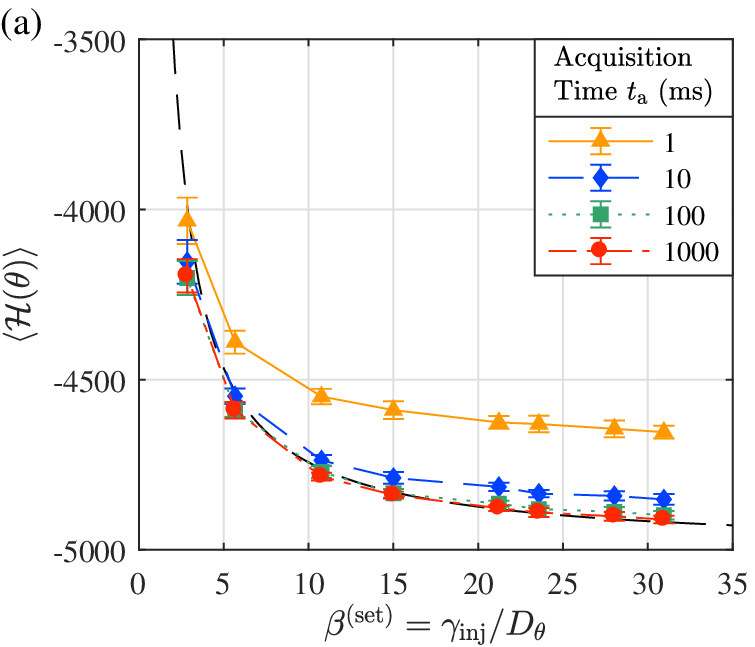}
  		\end{center}
 	\end{minipage}
 	\begin{minipage}{0.5\hsize}
  		\begin{center}
   		\includegraphics{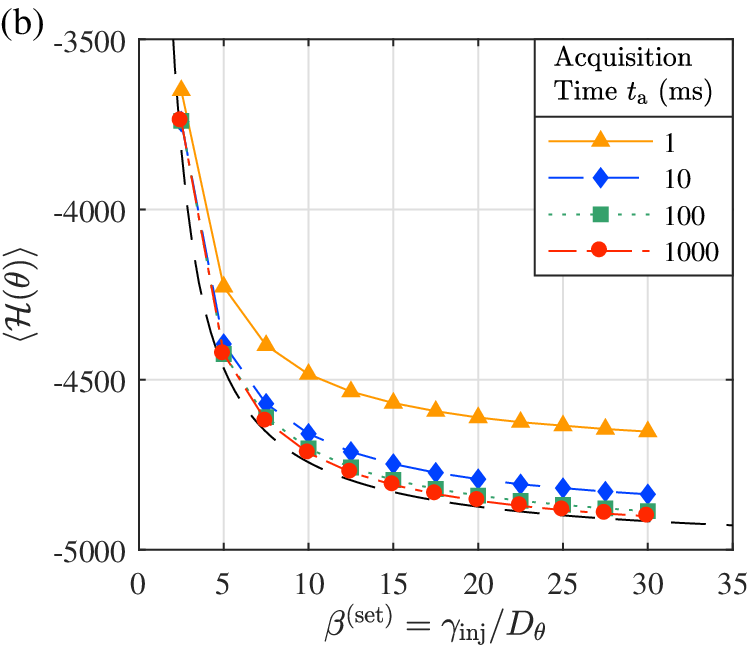}
  		\end{center}
 	\end{minipage}
\caption{Mean energy $\bracket{\mathcal{H}(\vct{\theta})}$ averaged over independently acquired samples for various settings $\beta\sur{(set)}$. (a) Mean energy values obtained from experiment. The conditions are the same as those for Figure \ref{figure5}(a). The error bars indicate standard deviations of the samples. (b) Mean energy values estimated through numerical simulations. The black dashed curve indicates the theoretical value (definition is given as $\bracket{\mathcal{H}(\vct{\theta})} = -N I_1(\beta)/I_0(\beta)$ \eref{eq:B5}, where $I_n$ is a modified Bessel function of the first kind).}
\label{figure6}
\end{figure}

\Fref{figure6}(a) plots the mean energy, $\bracket{\mathcal{H}(\vct{\theta})}$, which is averaged over the acquired samples. There were 1000 independent samples for $t\sub{a}=$1-100 ms and 100 independent samples for $t\sub{a}=$1000 ms.

\section{Discussion}
The measured effective temperature $\beta\sur{(eff)}$ that is shown in Figure \ref{figure5} increases, as the data acquisition time $t_\mathrm{a}$ becomes later. This trend is due to a relaxation process where neighboring XY spins change their directions to reduce the relative phases, and the total energy decreases as a time passes, as seen in Figure \ref{figure6}. The $\beta\sur{(eff)}$ converges on the setting values within $t_\mathrm{a} = $1000 ms for high temperature parameters $\gamma_\mathrm{inj}/D_\theta < 10$. When our system is used for real applications, these behaviors of converging time scales will determine the optimal parameter set under the trade-off between time and precision. 

We confirmed that the observed time scales were reasonable in terms of the Langevin dynamics. We compared the experimental results in Figure \ref{figure5}(a) with those estimated from the direct numerical simulations of the Langevin equations \eref{eq:Kuramoto} with the same $\gamma_\mathrm{inj}$ and $D_\theta$ values used in the experiment. \Fref{figure5}(b) shows the effective temperature $\beta\sur{(eff)}$ achieved by the numerical simulations. The convergence time scale of $\beta\sur{(eff)}$ to $\beta\sur{(set)}$ indicates good agreement between the experiment and the numerical simulations.

We evaluated the error on the setting values of interaction strength $J_{kl}$ and temperature parameter $\beta\sur{(set)}$. The steady state photon number should be the same for all NOPOs in our scheme to implement the Hamiltonian accurately. Otherwise, the photon number fluctuations would cause error on $J_{kl}$. We considered a situation where each photon number $n_k$ reached the steady state value determined by \eref{eq:9}, but each NOPO may have different photon number $n_k$. In that case, $J_{kl}$ and $\beta\sur{(set)}$ are modified as $J_{kl}\to(\sqrt{n_k n_l}/\bracket{n})J_{kl}$ and $\beta\sur{(set)}\to\gamma\sub{inj}\bracket{n}/D$, where $\bracket{n}$ is the ensemble-averaged photon number. The photon number fluctuation was estimated to be $\delta = 0.024$, where $\sqrt{n_k/\bracket{n}} = 1\pm\delta$ , through the amplitude fluctuation of the IQ signal in \fref{figure2}(b). Therefore the parameter noise for $J_{kl}$ due to photon number fluctuation was $2\delta = 4.8\%$ in our system, which was improved from $15\%$ in the previous work\cite{Tamate2016}.

The $\beta\sur{(set)}$ can be also affected by the ensemble-averaged photon number $\bracket{n}$. The achieved temperature $\beta\sur{(eff)}$ in \fref{figure5}(a) could have higher values than $\beta\sur{(eff)}$ estimated from numerical simulations (\fref{figure5}(a)). Moreover, some of the $\beta\sur{(eff)}$ in \fref{figure5}(a) exceeded their setting values $\beta\sur{(set)}$. These discrepancies imply that the actual $\beta\sur{(set)}$ turned to be set higher than the aimed values. Because we estimated $D_\theta$ from the differential phase decay without mutual couplings, the steady state photon number, $\bracket{n}$, possibly decreased, which overestimated $D_\theta$. This is one possible reason why the temperature parameter $\beta\sur{(set)}$ had an error on the lower side.

The achievable parameter range of $\beta\sur{(set)}$ in this system is limited by the mutual injection rate, $\gamma\sub{inj}$, and the phase diffusion coefficient, $D_\theta$. The maximum value of $\gamma\sub{inj}$ is $\sim$15 kHz, which is limited by coupling ratios of optical couplers in delay lines for mutual couplings. The minimum value of $D_\theta$ is $\sim$0.5 kHz, which is determined by intrinsic phase noise in the NOPO. Thus the achievable range is limited in $\beta\sur{(set)} < 30$, while $\beta\sur{(set)}$ can be set to extremely small values by reducing mutual injections and increasing noise injection. We need to increase $\gamma\sub{inj}$ and decrease $D_\theta$ to expand the range of $\beta\sur{(set)}$. The $\gamma\sub{inj}$ can be increased by changing the coupling ratios of the optical couplers in the delay lines or amplifying extracted fields before injecting them into other NOPOs. The $D_\theta$ can be decreased by reducing the cavity decay rate.

Controllability of the effective temperature is not only beneficial to the applications that require sampling at a designated temperature, but also to advanced research that applies techniques to manipulating temperatures to accelerate relaxation in Monte Carlo simulations, such as replica exchange MCMC sampling \cite{Hukushima1996}. The effective temperature was changed by one order of magnitude in this work, which was sufficient to implement these techniques and observe accelerations.

\section{Conclusion}
We experimentally studied the potential of the non-degenerate optical parametric oscillator (NOPO) network as a Boltzmann distribution sampler for a classical XY model. The NOPO network simulated the dynamics of a one-dimensional XY model with 5000 spins. The effective temperature as a Boltzmann distribution sampler was controlled via two experimental parameters, i.e., mutual coupling strength and external noise power. The experimental results indicated good agreement with numerical simulations of the Langevin equations, which confirmed that our scheme was correctly implemented. We hope that this work will motivate research in computation with physical systems and in algorithms that involve continuous and directional data.

\ack 
This research was funded by the Impulsing Paradigm Change through Disruptive Technologies (ImPACT) Program of the Council of Science, Technology and Innovation (Cabinet Office, Government of Japan). This work was supported by the RIKEN Center for emergent matter science.

\appendix
\section{Theory of one-dimensional XY ring}
\label{sec:theoryxy}
The statistical features of the one-dimensional XY model can be calculated by using the transfer matrix approach \cite{Tamate2016,Mattis1984}. The partition function of the one-dimensional XY ring with $J=1$ is expressed by:
\begin{eqnarray}
  Z = \sum_{n=-\infty}^{\infty} I_n(\beta)^N,  \label{eq:B1}
\end{eqnarray}
where $N$ is the number of spins and $I_n$ is the modified Bessel function of the first kind.

The probability distribution of the relative phase, $\tilde{\theta}$, between adjacent spins is given by:
\begin{eqnarray}
  p(\tilde{\theta}) = \frac{\exp(K\cos\tilde{\theta})}{2\pi Z}\sum_{n=-\infty}^{\infty} \cos(n\tilde{\theta}) I_n(\beta)^{N-1}.  \label{eq:B2}
\end{eqnarray}
The ensemble average of the energy is given by:
\begin{eqnarray}
  \bracket{\mathcal{H}(\vct{\theta})} = -\frac{N}{Z} \sum_{n=-\infty}^{\infty} I_n(K)^{N-1} I_{n+1}(\beta). \label{eq:B3}
\end{eqnarray}

When there are a sufficient number of spins, we can approximate Equations~\eref{eq:B2} and \eref{eq:B3} as:
\begin{eqnarray}
  p(\tilde{\theta}) = \frac{\exp(\beta \cos (\tilde{\theta}))}{2\pi I_0(\beta)}  \label{eq:B4}, \\
  \bracket{\mathcal{H}(\vct{\theta})} = -N\frac{I_1(\beta)}{I_0(\beta)}.\label{eq:B5}
\end{eqnarray}

\section{Phase measurements}
\label{sec:PM}
We measured two kinds of phases by means of coherent detection techniques in fiber-optic communication: the relative phases between two adjacent NOPOs and the individual phases of NOPOs in which the phase reference was an external cw laser. The relative phases were measured by using a one-interval delay in-phase/quadrature-phase (IQ) interferometer. The absolute phases were measured with a 90-degree optical hybrid with an external cw field for phase reference and balanced photo-detectors. The external cw laser used as a reference of the individual phase measurement was also employed for stabilizing the lengths of the cavity and the delay lines. There was frequency difference between the cavity longitudinal modes and the external cw laser because of XPM in HNLF. We compensated for the effect of the frequency difference, by acquiring the phases twice in series and exploiting the fact that the cavity round trip time was shorter than the time scale of phase diffusion. The details on these two methods of measurement are described in Tamate et al. \cite{Tamate2016}.

\section{Control of effective temperature by coupling strength between OPOs}
\label{sec:gamma_inj}
\begin{figure}[htbp]
 	\begin{minipage}{0.5\hsize}
  		\begin{center}
   		\includegraphics{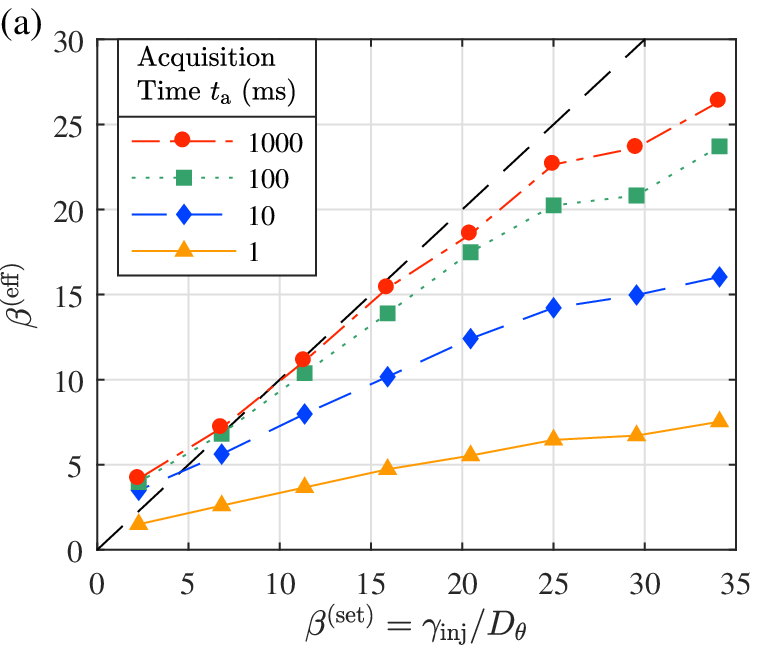}
  		\end{center}
 	\end{minipage}
 	\begin{minipage}{0.5\hsize}
  		\begin{center}
   		\includegraphics{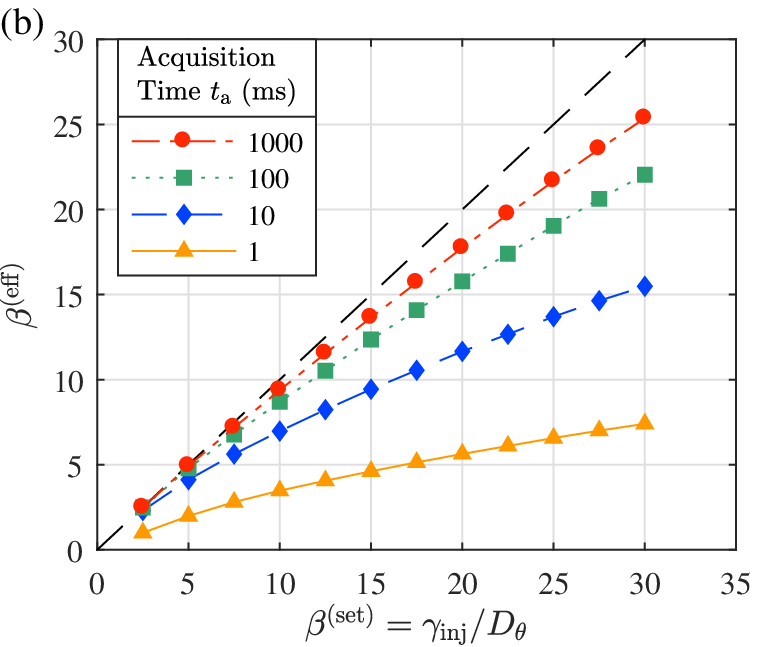}
  		\end{center}
 	\end{minipage}
\caption{Achieved effective temperature $\beta\sur{(eff)}$ for various settings $\beta\sur{(set)}$. The setting values of $\beta\sur{(set)}$ are varied by changing $\gamma_\mathrm{inj}$, while the phase diffusion coefficient was fixed, $D_\theta$ = 0.44 kHz. (a) $\beta\sur{(eff)}$ values obtained from experiment. (b) $\beta\sur{(eff)}$ values estimated through numerical simulations. Black dashed line indicates where $\beta\sur{(eff)} = \beta\sur{(set)}$.}
\label{figureC1}
\end{figure}

\begin{figure}[htbp]
 	\begin{minipage}{0.5\hsize}
  		\begin{center}
   		\includegraphics{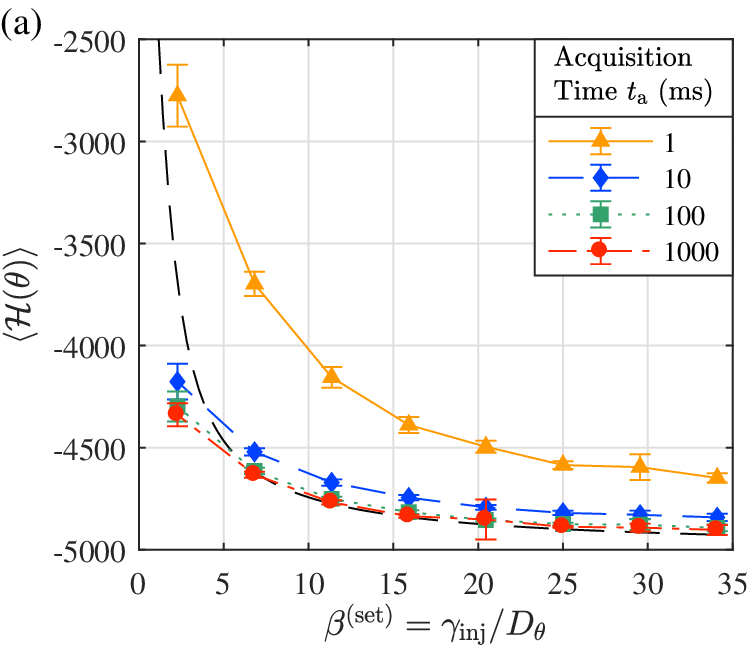}
  		\end{center}
 	\end{minipage}
 	\begin{minipage}{0.5\hsize}
  		\begin{center}
   		\includegraphics{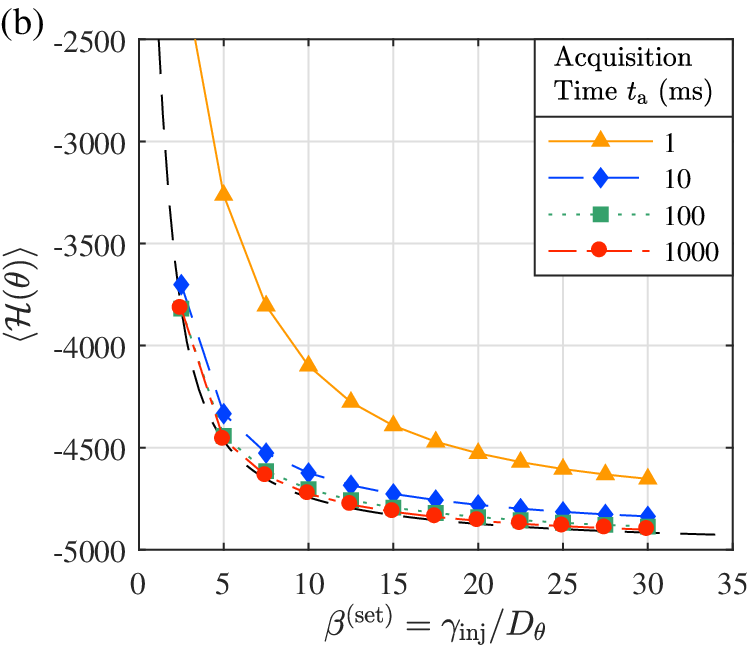}
  		\end{center}
 	\end{minipage}
\caption{Mean energy $\bracket{\mathcal{H}(\vct{\theta})}$ averaged over independently acquired samples for various settings $\beta\sur{(set)}$. (a) Mean energy values obtained from experiment. The conditions are the same as those for Figure\ref{figureC1}(a). The error bars indicate standard deviations for the samples. (b) Mean energy values estimated through numerical simulations. The black dashed curve indicates the theoretical value (definition is given as $\bracket{\mathcal{H}(\vct{\theta})} = -N I_1(\beta)/I_0(\beta)$ \eref{eq:B5}, where $I_n$ is the modified Bessel function of the first kind).}
\label{figureC2}
\end{figure}

temperature parameter $\beta\sur{(set)}$ was configured via the ratio between $\gamma_\mathrm{inj}$ and $D_\theta$, so $\beta\sur{(set)}$ could also be controlled by changing $\gamma_\mathrm{inj}$. The injection rate, $\gamma_\mathrm{inj}$, was controlled with the transmittance of delay lines for mutual injections. We calculated the transmittance value for required temperature $\beta\sur{(set)}$ under given $D_\theta$, and set these values with attenuators in delay lines. We fixed phase diffusionn coefficient, $D_\theta$ = 0.44 kHz, and varied $\gamma_\mathrm{inj}$ by tuning the transmittance, $T$, of optical delay lines for mutual couplings. Figure \ref{figureC1}(a), \ref{figureC2}(a) provides experimental results that correspond to the results in Figure \ref{figure5},\ref{figure6}. The difference between the cases in which $\beta\sur{(set)}$ is decreased by increasing $D_\theta$ and decreasing $\gamma_\mathrm{inj}$, is in the times scale of convergence to the setting value. Even for the same ratio, $\gamma_\mathrm{inj}/D_\theta$, the effective temperature $\beta\sur{(eff)}$ converges faster when $D_\theta$ is higher. This can also be seen in the numerical simulations in Figure \ref{figureC1}(b), \ref{figureC2}(b).

\end{document}